\documentclass[11pt]{article}

\usepackage[dvips]{graphicx}
\usepackage{amsfonts}

\newtheorem{remark}{Remark}

\newcommand{\R}{{\mathord{\mathbb R}}}
\newcommand{\C}{{\mathord{\mathbb C}}}

\newcommand{\beq}{\begin{equation}}
\newcommand{\eeq}{\end{equation}}

\newcommand{\br}{{\bf r}}
\newcommand{\bp}{{\bf p}}
\newcommand{\bq}{{\bf q}}
\newcommand{\by}{{\bf y}}
\newcommand{\bj}{{\bf j}}
\newcommand{\bv}{{\bf v}}

\newcommand{\bR}{{\bf R}}
\newcommand{\bL}{{\bf L}}
\newcommand{\bP}{{\bf P}}
\newcommand{\bA}{{\bf A}}
\newcommand{\bB}{{\bf B}}

\newcommand{\rmG}{{\rm G}}

\newcommand{\bcZ}{\mbox{\boldmath ${\cal Z}$}}
\newcommand{\CM}{{\mbox{\tiny C.M.}}}
\newcommand{\ST}{{\mbox{\tiny st}}}
\newcommand{\SP}{{\mbox{\tiny spin}}}
\newcommand{\SPT}{{\mbox{\tiny space}}}
\newcommand{\IS}{{\mbox{\tiny isospin}}}

\newcommand{\bxi}{\mbox{\boldmath $\xi$}}

\newcommand{\subbxi}{\mbox{\boldmath $\scriptstyle \xi$}}

\newcommand{\bXi}{{\bf \Xi}}
\newcommand{\bPi}{{\bf \Pi}}

\newcommand{\inter}{{({\rm i})}}

\newcommand{\gs}{{|0\!\! >}}

\newcommand{\Imag}{{\mbox{Im}\,}}
\newcommand{\SImag}{{\mbox{\small Im}\,}}
\newcommand{\Real}{{\mbox{Re}\,}}

\newcommand{\Cons}{\, \mbox{Const.} \,}

\newcommand{\sds}{\strut\displaystyle}
\newcommand{\smallspace}{\vspace{2.5ex}}

\begin{document}

\title{\bf{Non--Local Potentials and Rotational Bands of Resonances in Ion Collisions}}

\author{\vspace{5pt} Enrico De Micheli$^{\dag}$ and Giovanni Alberto Viano$^{\ddag}$ \\
$^{\dag}$\small{Istituto di Cibernetica e Biofisica -- Consiglio Nazionale delle Ricerche}\\[-5pt]
\small{Via De Marini, 6 -- 16149 Genova, Italy. \vspace{8pt}} \\
$^{\ddag}$\small{Dipartimento di Fisica -- Universit\`a di Genova} \\[-5pt]
\small{Istituto Nazionale di Fisica Nucleare -- sez. di Genova} \\[-5pt]
\small{Via Dodecaneso, 33 -- 16146 Genova, Italy.}
}

\date{}

\maketitle

\begin{abstract}
Sequences of rotational resonances (rotational bands)
and corresponding antiresonances are observed in ion collisions. 
In this paper we propose a description which
combines collective and single--particle features of
cluster collisions. It is shown how rotational bands emerge in many--body dynamics, when the
degeneracies proper of the harmonic oscillator spectra are removed by adding interactions depending on
the angular momentum.
These interactions can be properly introduced in connection with the exchange--forces and the antisymmetrization,
and give rise to a class of non--local potentials whose spectral properties are analyzed in detail.
In particular, we give a classification of the singularities of the resolvent, which are associated
with bound states and resonances. The latter are then studied using an appropriate type of
collective coordinates, and a hydrodynamical model of the trapping, responsible for the resonances,
is then proposed. Accordingly, we derive, from the uncertainty principle, a {\it spin--width} of the
unstable states which can be related to their angular lifetime.
\end{abstract} 

\newpage

\section{Introduction}
\label{se:introduction}
The phase--shifts $\delta_\ell$ in the $\alpha$--nucleus elastic scattering have,
at low energy, the following behaviour: first, they rise passing through $\pi/2$,
and cause a sharp maximum in the energy dependence of $\sin^2\delta_\ell$ that corresponds to a 
resonance; then they decrease, crossing $\pi/2$ downward, and produce what is
called an {\it echo} or an antiresonance \cite{McVoy1,McVoy2}. Typical examples are
the phase--shifts in the $\alpha$--$\alpha$ or in $\alpha$--$^{40}$Ca elastic scattering
(see refs. \cite{Okai,Langanke,Langanke2}). A second relevant feature is that the
resonances and, correspondingly, the antiresonances, are organized in an ordered sequence and produce a rotational
band of levels whose energy spectrum can be fitted by an expression of the form
$E_\ell = A+B \, \ell(\ell+1)$, where $\ell$ is the (approximate) orbital angular momentum of the level,
and $A$ and $B$ are constants almost independent of $\ell$. 
Finally, the widths of the resonances increase as a function of the energy; consequently, at higher energy,
the rotational resonances evolve into surface waves that can be explained as due to the
diffraction by the target, which is regarded as an opaque (or partially opaque) obstacle.
These features, which emerge in a particularly clear form in the $\alpha$--nucleus scattering
are however characteristic of a very large class of heavy ion collisions,
like {\it e.g.}, $^{12}$C--$^{12}$C, $^{12}$C--$^{16}$O, $^{24}$Mg--$^{24}$Mg, $^{28}$Si--$^{28}$Si, 
etc. (see \cite{Cindro} and the references quoted therein).

Though rotational resonances can be described, with some approximation, 
within the framework of the two--body problem, antiresonances necessarily involve the many--body 
properties of the interaction. In fact, the rotational levels can be regarded as produced by 
the trapping of the incoming projectile which rotates, for a certain time, around the target; 
thus, we have the energy spectrum of a rigid rotator. 
Conversely, the two--body model is totally inadequate for explaining the echoes. 
In this case, the two interacting particles ({\it e.g.}, $\alpha$ particles or
$^{40}$Ca ions) cannot simply be treated as bosons, because the wavefunction of the whole system 
must be antisymmetric with respect to the exchange of all the nucleons, including those belonging 
to different clusters.
Then, the fermionic character of the nucleons emerges, and from the antisymmetrization a 
repulsive force derives; accordingly, the phase shifts $\delta_\ell$ decrease, and antiresonances are produced.
We are thus led to combine the collective and the single--particle features of cluster collisions.

In section \ref{se:algebraic} the many--body problem, treated by means of the Jacobi coordinates, 
will be briefly sketched. By introducing potentials of harmonic oscillator type, it can be observed 
the formation of clusters whose ground--state wavefunction can be represented by the product of gaussians.
Furthermore, if the degeneracies, proper of the harmonic oscillator, are removed
by introducing interactions which depend on the angular momentum, then rotational bands emerge from
the many--body dynamics.

In section \ref{se:many} the relationship between the Jacobi coordinates and the relative coordinates
of the interacting clusters is addressed. We are thus led to consider the wavefunction which describes the relative 
motion of the clusters.
In particular, the antisymmetrization and the exchange character of the nuclear forces yield non--local
potentials. To this end, it is worth remembering that the derivation of non--local
potentials from many--body dynamics has been extensively studied in the past \cite{Wildermuth1}.
The procedures commonly used are:
the resonating group method \cite{Tang}, the complex generator coordinate technique \cite{LeMere},
and the cluster coordinate method \cite{Wildermuth2}. Their common goal is to provide a 
microscopic description of the nuclear processes that, starting from a nucleon--nucleon potential 
and employing totally antisymmetric wavefunctions, can evaluate bound states and scattering cross sections 
from an unified viewpoint.
All these methods have been remarkably successful at a computational level, and they will not be considered
hereafter since our main interest here concerns the spectral analysis associated with non--local potentials. 
The mathematical tool used in this connection is the Fredholm alternative \cite{Riesz}, which allows us to 
study the analytical properties of the
resolvent associated with the integro--differential equation of the relative motion, and the main properties
of bound states, resonances and antiresonances generated by an appropriate class of non--local potentials.
Particular attention is devoted to the fact that the non--local potentials represent angular--momentum dependent
interactions and, therefore, in view of the results obtained in section \ref{se:algebraic}, 
they can appropriately describe rotational bands.

Section \ref{se:collective} is devoted to the rotational resonances regarded as a collective
phenomenon. To this end, we introduce a type of collective coordinates (called $Z$--coordinates)
and first analyze the relationship between Jacobi and $Z$--coordinates, and then between
the latter and the relative coordinates of the interacting clusters. Then, in this scheme, 
we can introduce a hydrodynamical model of the trapping which is able to {\it generate the resonances}.
Finally, from this model and through the uncertainty principle, a
{\it spin--width}, which is proper of rotational resonances, will be defined.

In this paper we focus on the following questions
which we think have not received enough attention so far:
\begin{itemize}
\item[i)] An algebraic--geometric analysis which shows how rotational bands emerge from the
many--body dynamics (see section \ref{se:algebraic}).
\item[ii)] The spectral analysis associated with non-local potentials that can give a classification
of the resolvent singularities corresponding to bound states and resonances (see section \ref{se:many}).
\item[iii)] The relationship between Jacobi and $Z$--coordinates, which allows us to describe the collective
features of the resonances, and to introduce the spin--width proper of the resonances 
(see section \ref{se:collective}).
\end{itemize}

\section{An Outline of the Algebraic--Geometric Approach to Rotational Bands}
\label{se:algebraic}
For the reader's convenience, in this section we briefly review the Jacobi
and the hyperspherical coordinates (very well known in the literature \cite{Dragt,Simonov,Badalayan}), 
focusing on those algebraic--geometric aspects which play a relevant role in the following.

First, let us consider the case of three particles of equal mass $m$, whose positions
are described by the vectors $\br_k = (x_k,y_k,z_k)$, $(k=1,2,3)$. The kinetic energy operator reads
\beq
\label{uno}
T = -\frac{1}{2m}\left (\Delta_1+\Delta_2+\Delta_3 \right ),~~~~~(\hbar = 1)\,,
\eeq
where $\Delta_k :\equiv \partial^2/\partial x_k^2 + \partial^2/\partial y_k^2 + \partial^2/\partial z_k^2$,
$(k=1,2,3)$. We can now introduce the Jacobi and center of mass coordinates, which are defined as follows:
\begin{eqnarray}
\sds \bxi_1 &=& \frac{\br_1-\br_2}{\sqrt{2}}, \label{duea}\\
\sds \bxi_2 &=& \left (\frac{2}{3}\right )^{1/2} \left (\frac{\br_1+\br_2}{2}-\br_3\right ), \label{dueb} \\
\sds \bR_\CM &=& \frac{\br_1+\br_2+\br_3}{3}~,~~~~~|\br_k|=\sqrt{x_k^2+y_k^2+z_k^2}\,. \label{duec}
\end{eqnarray}
The kinetic energy operator can be written in terms of these coordinates as
\beq
\label{tre}
T=-\frac{1}{2m}\left\{\Delta_{{\subbxi}_1} + \Delta_{{\subbxi}_2} + \frac{1}{3}\Delta_{{\bR}_\CM} \right\},
\eeq
where
\begin{eqnarray}
\Delta_{{\subbxi}_i} \equiv \frac{\partial^2}{[\partial (\bxi_i)_x]^2} + 
\frac{\partial^2}{[\partial (\bxi_i)_y]^2} + \frac{\partial^2}{[\partial (\bxi_i)_z]^2},~~(i=1,2),~~~\\
\Delta_{{\bR}_\CM} \equiv \frac{\partial^2}{[\partial (\bR_\CM)_x]^2} + 
\frac{\partial^2}{[\partial (\bR_\CM)_y]^2} + \frac{\partial^2}{[\partial (\bR_\CM)_z]^2},~~
\end{eqnarray}
$(\bxi_i)_x, (\bxi_i)_y, (\bxi_i)_z$ and $(\bR_\CM)_x, (\bR_\CM)_y, (\bR_\CM)_z$ denoting
the $x,y,z$ components of the vectors $\bxi_i$ and $\bR_\CM$, respectively. 
Then, the kinetic energy of the center of mass can be separated from that of the relative motion $T_R$:
\beq
\label{quattro}
T_R = -\frac{1}{2m} \left\{\Delta_{\subbxi_1}+\Delta_{\subbxi_2} \right\}.
\eeq
Now, it is convenient to combine the vectors $\bxi_1$ and $\bxi_2$ into a single vector
${\bXi} = {\subbxi_1 \choose \subbxi_2}$, whose Cartesian components
will be denoted by $\bXi_1, \bXi_2,\ldots,\bXi_6$.
Thus, we can consider a sphere embedded in $\R^6$ whose radius is $\rho^2=\xi_1^2+\xi_2^2$, and,
accordingly, represent the components of $\bXi$
in terms of the spherical coordinates $(\rho, \theta_1,\ldots,\theta_5)$ as follows:
\begin{eqnarray}
\Xi_1 & = & \rho\sin\theta_5\sin\theta_4\cdots\sin\theta_1, \nonumber \\
\Xi_2 & = & \rho\sin\theta_5\sin\theta_4\cdots\cos\theta_1, \nonumber \\
\cdots & & \cdots\cdots \\
\Xi_5 & = & \rho\sin\theta_5\cos\theta_4, \nonumber \\
\Xi_6 & = & \rho\cos\theta_5.\nonumber 
\end{eqnarray}
In terms of spherical coordinates the Laplace--Beltrami operator $\Delta$ reads \cite{Vilenkin}
\begin{eqnarray}
\label{cinque}
\Delta &=& \frac{1}{\rho^5}\frac{\partial}{\partial\rho}\left (\rho^5\frac{\partial}{\partial\rho}\right ) +
\frac{1}{\rho^2\sin^4\theta_5}\frac{\partial}{\partial\theta_5}
\left (\sin^4\theta_5\frac{\partial}{\partial\theta_5}\right )
+\frac{1}{\rho^2\sin^2\theta_5\sin^3\theta_4}\frac{\partial}{\partial\theta_4}
\left (\sin^3\theta_4\frac{\partial}{\partial\theta_4}\right ) + \cdots \nonumber \\
&+&\frac{1}{\rho^2\sin^2\theta_5\sin^2\theta_4\cdots\sin^2\theta_2}\frac{\partial^2}{\partial\theta_1^2},
\end{eqnarray}
and, by separating the radial part from the angular one, we get:
\beq
\label{sei}
\Delta=\frac{1}{\rho^5}\frac{\partial}{\partial\rho}\left (\rho^5\frac{\partial}{\partial\rho}\right ) +
\frac{1}{\rho^2}\Delta_0,
\eeq
where $\Delta_0$ is the Laplace--Beltrami operator acting on the unit sphere ${\cal S}^5$
embedded in $\R^6$ \cite{Vilenkin}. \\
Let us introduce the harmonic polynomials of degree $j$ \cite{Vilenkin}, which may be written as
$\rho^j\Theta_j(\theta_1,\ldots\theta_5)$. Then, from (\ref{sei}) we get:
\beq
\label{sette}
\Delta\left [\rho^j\Theta_j(\theta_1,\ldots\theta_5)\right]=j(j+4)\rho^{(j-2)}\Theta_j(\theta_1,\ldots\theta_5)
+\rho^{(j-2)}\Delta_0\Theta_j(\theta_1,\ldots\theta_5)=0,
\eeq
which gives
\beq
\label{otto}
\Delta_0\Theta_j(\theta_1,\ldots\theta_5)=-j(j+4)\Theta_j(\theta_1,\ldots\theta_5).
\eeq
Next, we introduce the momenta associated with the Jacobi coordinates, {\it i.e.},
\begin{eqnarray}
&&\sds \bp_{\xi_1} = \frac{\bq_1-\bq_2}{\sqrt{2}}, \label{novea}\\
&&\sds \bp_{\xi_2} = \left (\frac{2}{3}\right )^{1/2} \left (\frac{\bq_1+\bq_2}{2}-\bq_3\right ), \label{noveb}
\end{eqnarray}
where $(\bq_k)_x = m\dot{x}_k,\, (\bq_k)_y = m\dot{y}_k,\, (\bq_k)_z = m\dot{z}_k$ $(k=1,2,3)$, and also
combine the momenta in a single vector $\bP = {\bp_{\xi_1} \choose \bp_{\xi_2}}$.
Then, we consider a potential of the form
\beq
\label{quattordici}
V(\rho)=\rmG\left\{|\br_1-\br_2|^2+|\br_1-\br_3|^2+|\br_2-\br_3|^2\right\}=3\rmG\rho^2.
\eeq
Again, by separating in the wavefunction $\psi(\rho;\theta_1,\ldots,\theta_5)$ the radial variable 
from the angular ones, we have the following equations:
\begin{eqnarray}
&&\sds \frac{1}{\rho^5}\frac{d}{d\rho}\left (\rho^5\frac{dR_j}{d\rho}\right )-\frac{j(j+4)}{\rho^2}R_j
+ 2m\left [ E-V(\rho)\right ] R_j = 0, \label{quindicia}\\
&&\sds \Delta_0\Theta_j(\theta_1,\ldots\theta_5)=-j(j+4)\Theta_j(\theta_1,\ldots\theta_5), \label{quindicib}
\end{eqnarray}
where $E$ denotes the energy. It is easy to see that the solutions of eq. (\ref{quindicia}) are given by
\begin{eqnarray}
&& R_j(\rho) = \rho^j \exp\left [-\frac{1}{2}\sigma^2\rho^2\right ],~~~~~\sigma=\left (mK\right )^{1/4},
\label{sedicia}\\
&&\sds E_j = (j+3)\,\omega,~~~~~\omega=\left (\frac{K}{m}\right )^{1/2}, \label{sedicib}
\end{eqnarray}
where $K=6\rmG$.

It is well known that the group of the permutations of three objects has two one--dimensional representations
and one two--dimensional representation. A remarkable fact is that the elements of the permutation group 
lead to rotations in $\R^6$, but, as stated in \cite{Dragt}, not all the elements 
of the Lie algebra associated with the $SO(6)$ group treat the three particles equivalently. 
On the other hand, the sphere ${\cal S}^5$ may be regarded as the unit sphere embedded in $\C^3$ 
because the complex vector 
space $\C^3$ can be identified with the space $\R^{6}$.
Therefore ${\cal S}^5$ may be identified with $SU(3)/SU(2)$, $SU(3)$ acting transitively on
${\cal S}^5$. Thus, we are naturally led to introduce the complex vectors:
\begin{eqnarray}
\bcZ = \bxi_1 + i \bxi_2, & ~~~~~~~~~~~ & \bcZ^* = \bxi_1 - i \bxi_2, \label{diciassettea} \\
\bPi = \bp_{\xi_1} + i \bp_{\xi_2}, & ~~~~~~~~~~~ & \bPi^* = \bp_{\xi_1} - i \bp_{\xi_2}, \label{diciassetteb}
\end{eqnarray}
and to reformulate the problem in the $SU(3)$ group framework. 
It is easy to see \cite{Simonov} that the operators of interchange of particles turn
$\bcZ$ into $\bcZ^*$, and vice--versa, with multiplication by a complex number. 
Then, we have:
\begin{eqnarray}
&& \bcZ\cdot\bcZ^* = \xi_1^2 + \xi_2^2 = \rho^2, \label{diciannovea} \\
&& \bPi\cdot\bPi^* = p_{\xi_1}^2 + p_{\xi_2}^2 = -\Delta.  \label{diciannoveb}
\end{eqnarray}
Next, if we set $m=\hbar=1$, $\rmG=1/6$, $(K=1)$,
the total Hamiltonian can be written in the following form:
\beq
\label{venti}
H = -\frac{1}{2}\Delta+V=\frac{1}{2}\left (\bPi\cdot\bPi^*+\bcZ\cdot\bcZ^*\right ).
\eeq
Now, in order to deal with the harmonic oscillator problem in the Fock space, we introduce
the vector creation and annihilation operators \cite{Dragt}:
\begin{eqnarray}
&&\bA^\dagger = \frac{1}{\sqrt{2}}\left (\bxi_1-i\bp_{\xi_1}\right ),~~~
  \bA=\frac{1}{\sqrt{2}}\left (\bxi_1+i\bp_{\xi_1}\right ), \label{ventunoa} \\
&&\bB^\dagger = \frac{1}{\sqrt{2}}\left (\bxi_2-i\bp_{\xi_2}\right ),~~~
  \bB=\frac{1}{\sqrt{2}}\left (\bxi_2+i\bp_{\xi_2}\right ), \label{ventunob}
\end{eqnarray}
which satisfy the following commutation rules
\begin{eqnarray}
&&\left [A_k,A^\dagger_p\right ] = \delta_{kp},~~~~~~(k,p=1,2,3), \label{ventiduea} \\
&& \left [B_k,B^\dagger_p\right ] = \delta_{kp}.\label{ventidueb}
\end{eqnarray}
Finally, in view of the commutation rules (\ref{ventiduea}, \ref{ventidueb}), the Hamiltonian can be rewritten 
as follows:
\beq
\label{venticinque}
H=\bA^\dagger\cdot\bA+\bB^\dagger\cdot\bB+3=N_A+N_B+3,
\eeq
where $N_A$ and $N_B$ are the occupation numbers associated with the operators $\bA^\dagger\cdot\bA$ and
$\bB^\dagger\cdot\bB$, respectively. Then, for the ground state $\gs$, which is characterized by the conditions
$\bA\gs=\bB\gs=0$, we have $H\gs=3\gs$, which represents the zero point energy;
correspondingly, the wavefunction is given by: $\exp[-\rho^2/2]$.

Let $j_1$ and $j_2$ denote the eigenvalues of $N_A$ and $N_B$, respectively. Then, from
(\ref{sedicib}) and (\ref{venticinque}), we get: $j=j_1+j_2$.
As is well known, from the Cartan analysis for the $SU(3)$ group, any irreducible representation
of $SU(3)$ is completely characterized by two indexes which, in our case, are precisely $j_1$
and $j_2$ (for the Cartan classification indexes see refs. \cite{Dragt,Hamermesh,Weyl}). 
In order to investigate how rotational sequences emerge from the three--body dynamics, 
first the total angular momentum $\bL$ about the center of mass must be introduced, {\it i.e.},
\beq
\label{ventisette}
\bL = \br_1\times\bq_1+\br_2\times\bq_2+\br_3\times\bq_3-\bR_{C.M.}\times\bp_{R_{C.M.}},
\eeq
(where $\bp_{R_{C.M.}}=(\bq_1+\bq_2+\bq_3)/3)$, which may also be interpreted as the total angular momentum
in the center--of--momentum frame, since in this case $\bp_{R_{C.M.}}=0$. Then we have to answer to the 
following question:

\smallspace
\noindent
\underline{Problem}:
Determine the $\ell$--values, $\ell(\ell+1)$ being the eigenvalues of ${\bf L}^2$,
contained in the representation $(j_1,j_2)$, where $j_1$ and $j_2$ are the Cartan indexes of $SU(3)$.

\smallspace
This problem may be rephrased as follows: determine what irreducible representations of the group $SO(3)$, 
which are labelled by $\ell$, occur in an irreducible representation of the group $SU(3)$.
Following Weyl \cite{Weyl}, the irreducible representations of the group $U(n)$ are labelled by 
a set of non--negative integers $f_1,f_2,\ldots,f_n$ such that: $f_1\geq f_2 \geq \cdots \geq f_n$.
Next, in the reduction of $U(n)$ to $SU(n)$ the irreducible representations
$(f_1,f_2,\ldots,f_n)$ of $U(n)$ remain irreducible under $SU(n)$, but a simplification
occurs since certain representations which are not equivalent under $U(n)$ become equivalent under $SU(n)$.
Precisely, $(f_1,f_2,\ldots,f_n)$ become equivalent to $(f_1-f_n,\ldots,f_{n-1}-f_n)$. Consequently, for $SU(3)$,
the partition $(f_1,f_2,f_3)$ can be replaced by the differences $k_1=f_1-f_3,k_2=f_2-f_3$, which
can be related to $j_1$ and $j_2$ as follows \cite{Bargmann}:
\beq
k_1=j_1+j_2\,,~~~~~~ k_2 = j_1. \label{ventotto}
\eeq
Furthermore, the Weyl approach gives the expression of the characters, and in particular the
dimension, of the irreducible representations of the $U(n)$ or $SU(n)$ group. The formula for the dimension reads
\cite{Weyl}
\beq
\label{ventinove}
\mbox{dim}\, (f_1,f_2,\ldots,f_n)=\prod_{1\leq i < k \leq n}\left (\frac{f_i-f_k+k-i}{k-i}\right ).
\eeq
Therefore, we answer the question posed by the problem formulated above by means of the following 
equality:
\beq
\label{trenta}
\prod_{1\leq i < k \leq 3}\left (\frac{f_i-f_k+k-i}{k-i}\right )=\sum_\ell\mu_\ell (2\ell+1),
\eeq
where $(2\ell+1)$ is the dimension of the representation $D_\ell$ of the rotation group, while $\mu_\ell$
denotes the multiplicity, {\it i.e.}, it gives the number of times the representation $D_\ell$ occurs in a
certain representation of $SU(3)$.
Formula (\ref{trenta}) has been obtained by equating the characters of the representation
in the specific case of the unit element.

For convenience, we shall continue to use: $j_1=k_2=f_2-f_3$ and $j_2=k_1-k_2=f_1-f_2$. Then,
in our case, formula (\ref{trenta}) reads
\beq
\label{trentuno}
(j_1+1)(j_2+1) \left (\frac{j_1+j_2+2}{2}\right )=\sum_\ell\mu_\ell (2\ell+1).
\eeq
Since the l.h.s of (\ref{trentuno}) is symmetric in $j_1$ and $j_2$, it follows that
$\mbox{dim}\, (j_1,j_2)=\mbox{dim}\, (j_2,j_1)$. 
Now, we consider two cases:
\begin{itemize}
\item[a)] Let $j_1=2n$ ($n$ integer) and $j_2=0$; then:
\beq
\label{trentadue}
\mbox{dim}\, (j_1,j_2)=\mbox{dim}\, (2n,0)=(n+1)(2n+1)
=\mbox{dim}\, (D_0+D_2+\cdots +D_{2n}).
\eeq
This means that the $\ell$--values that occur in the representation $(2n,0)$ are: $\ell=0,2,4,\ldots,2n$;
$(\mu_\ell=1)$. We have thus obtained a rotational band of even parity.
\item[b)] Let $j_1=2n+1$ ($n$ integer) and $j_2=0$; then:
\beq
\label{trentaduebis}
\mbox{dim}\, (j_1,j_2)=\mbox{dim}\, (2n+1,0)=(n+1)(2n+1)
=\mbox{dim}\, (D_1+D_3+\cdots +D_{2n+1}).
\eeq
This means that the $\ell$--values occuring in the representation $(2n+1,0)$ are: $\ell=1,3,5,\ldots,2n+1$;
$(\mu_\ell=1)$. We have thus obtained a rotational band of odd parity.
\end{itemize}
\begin{remark} 
\rm
The physics of the harmonic oscillator from the group theoretical viewpoint has been
thoroughly investigated particularly by Moshinsky and his school (see, in this respect, the excellent
book by Moshinsky and Smirnov \cite{Moshinsky} and the references quoted therein). In particular,
the rule that emerges from formula (\ref{trenta}) is known in nuclear physics as the {\it Elliott rule}
\cite{Elliott1}, and it has been extensively used in connection with nuclear models and,
specifically, in the analysis of rotational and shell models \cite{Elliott2};
however, as far as we know, it has never been derived and used in the Jacobi approach to the many--body problem.
\end{remark}

Let us observe that levels with different values of $\ell$, but with the same value of $j=j_1+j_2$, are degenerate.
In order to remove this degeneracy, other interactions must be added to the harmonic oscillator Hamiltonian.
If a term proportional to $\bL\cdot\bL$ is added, we shall have a splitting within the $SU(3)$ multiplets, 
which is proportional to the square of the total angular momentum $\bL$. 
Since an energy spectrum proportional to $\bL^2$ is just a rotational spectrum,
we see that each $SU(3)$ multiplet gives rise to a rotational band. {\it The analysis indicates that rotational
bands emerge from the three--body dynamics if angular momentum dependent interactions are acting}.

It is straightforward to generalize the Jacobi coordinates to $N$ identical particles
of mass $m$. We have:
\begin{eqnarray}
\bxi_1 & = & \frac{1}{\sqrt{2}} \left ( \br_1-\br_2 \right ), \nonumber \\
\cdots &   & \cdots\cdots \\
\bxi_{N-1} & = & \frac{1}{\{N(N-1)\}^{1/2}}\left (\sum_{n=1}^{N-1} \br_n-(N-1)\br_N\right ).
\nonumber
\end{eqnarray}
Then, we may introduce the hypersphere with radius $\rho$ given by
\beq
\label{trentatre}
\rho^2 = \frac{1}{2N}\sum_{k,p=1}^N |\br_k-\br_p|^2 = \bxi_1^2 + \bxi_2^2 + \cdots + \bxi_{N-1}^2.
\eeq
The kinetic energy operator for the $N$--body problem is
\beq
\label{trentaquattro}
T = -\frac{1}{2m}\left ( \Delta_1 + \Delta_2 + \cdots + \Delta_N \right ),~~~~~(\hbar = 1),
\eeq
where $\Delta_k=\partial^2/\partial x_k^2 + \partial^2/\partial y_k^2 + \partial^2/\partial z_k^2$,
$(k=1,2,\ldots,N)$. Through the Jacobi coordinates, the center of mass kinetic
energy can be separated from the kinetic energy of the relative motion $T_R$, which reads
\beq
\label{trentacinque}
T_R = -\frac{1}{2m}\left ( \Delta_{\subbxi_1} + \Delta_{\subbxi_2} + \cdots + \Delta_{\subbxi_{N-1}} \right ),
\eeq
where $\Delta_{\subbxi_i}=\partial^2/(\partial(\bxi_i)_x)^2 + \partial^2/(\partial(\bxi_i)_y)^2
+ \partial^2/(\partial(\bxi_i)_z)^2$, $(i=1,2,\ldots,N-1)$.
If we introduce the hyperspherical coordinates $(\rho,\theta_1,\theta_2,\ldots,\theta_{3N-4})$, and consider
a harmonic oscillator potential of the form:
\beq
\label{trentasei}
V(\rho)=\frac{1}{2} K \rho^2,
\eeq
we can write the Sch\"{o}dinger equation as follows:
\beq
\label{trentasette}
H\psi = \left ( -\frac{1}{2m} \Delta + V \right ) \psi\,,
\eeq
where $\Delta$ is the Laplace--Beltrami operator on the hypersphere of radius $\rho$. By separating
the radial variable from the angular ones, we are led to the following expression for the radial part 
of the wavefunction:
\beq
\label{trentotto}
R_j(\rho) = \rho^j \exp\,\left(-\frac{1}{2}\sigma^2\rho^2\right),
\eeq
where $\sigma^2 = m\omega$ $(\omega = \sqrt{K/m})$. Therefore, the radial part of the ground state wavefunction
reads
\beq
\label{trentanove}
R(\rho) = \exp\left(-\frac{m\omega}{2}\rho^2\right) = \exp\left(-\frac{m\omega}{2}\sum_{i=1}^{N-1}\xi_i^2\right).
\eeq
We can conclude that for any cluster of identical particles the radial wavefunctions of the ground state
can be brought to a product of gaussians of the form (\ref{trentanove}), if the potential has the
harmonic oscillator form (\ref{trentasei}). Finally, we can regard the constant $K$ as a degree of 
{\it compactness} of the cluster, by noting that the peak of the bell--shaped curve representing the function
$\exp (-m\omega\rho^2/2)$ becomes sharper for increasing values of $K$.

\section{Non--Local Potentials and the Associated Spectral Analysis}
\label{se:many}
From the analysis of the previous section we deduce that:
\begin{itemize}
\item[a)] In order to obtain rotational bands, forces that depend on the angular momentum must
be added to the harmonic oscillator potential. In this case a spectrum proportional to $\bL ^2$ is
produced.
\item[b)] Harmonic oscillator potentials can produce clusters of particles, whose compactness
is related to the force constant $K$, and whose radial wavefunctions are expressed as a product of gaussians.
\end{itemize}

On the other hand the phenomenology shows that rotational bands of resonances emerge from the collisions
of clusters. Therefore, in view of point (a), we could try to ascertain if and how these sequences of rotational
resonances emerge when interactions, which depend on the angular momentum,
are added to harmonic oscillator type forces.
In this section we only consider the one--channel case: the elastic channel in the scattering theory.

First, we introduce suitable coordinates for describing cluster collisions. 
Let $\bR_i$ $(i=1,2)$ be the center of mass coordinates of the two clusters. 
Then assuming, for the sake of simplicity, that the two clusters have the same number $n$ of nucleons, we have:
\begin{eqnarray}
\bR_1 & = & \frac{\br_1+\br_2+ \cdots + \br_n}{n}, \label{manyunoa} \\
\bR_2 & = & \frac{\br_{n+1}+\br_{n+2} + \cdots + \br_{2n}}{n}, \label{manyunob}
\end{eqnarray}
where $\br_j$ is the space coordinate of the $j$--th particle, and $2n=N$. The center of mass and
the relative coordinates of the two clusters are respectively given by:
\begin{eqnarray}
\bR_\CM & = & \frac{1}{2}(\bR_1+\bR_2), \label{manyduea} \\
\bR & = & \bR_1 - \bR_2\,. \label{manydueb}
\end{eqnarray}
Next, we introduce the following internal cluster coordinates $\br_j^\inter$ \cite{Wildermuth2}:
\begin{eqnarray}
\br_j^\inter & = & \br_j - \bR_1\,,~~~\mbox{if}~~j=1,2,\ldots,n, \label{manytrea} \\
\br_j^\inter & = & \br_j - \bR_2\,,~~~\mbox{if}~~j=(n+1),(n+2),\ldots,2n. ~~~~~\label{manytreb}
\end{eqnarray}
It can easily be verified that the following equality holds true:
\beq
\label{manyquattro}
\rho^2 = \sum_{j=1}^{n-1} \xi_j^2 = \sum_{j=1}^n \left(r_j^\inter\right)^2.
\eeq
Therefore, for each cluster, the spatial part of the ground state wavefunction can be rewritten in
terms of internal cluster coordinates as follows (see (\ref{trentanove})):
\begin{eqnarray}
&&\Phi^0_\SPT(1) = \exp\left(-\frac{m\omega}{2}\sum_{j=1}^n\left(r_j^\inter\right)^2\right),
\label{manycinquea} \\
&&\Phi^0_\SPT(2) = \exp\left(-\frac{m\omega}{2}\sum_{j=n+1}^{2n}\left(r_j^\inter\right)^2\right).
\label{manycinqueb}
\end{eqnarray}
The wavefunction that describes the system composed of the interacting clusters must be antisymmetric
with respect to the exchange of all the nucleons, including those belonging to different clusters.
Then the wavefunction of the system composed by the two clusters is \cite{Wildermuth2}
%\begin{eqnarray}
%\label{manysei}
%\Psi& = & {\cal A}\left\{\left[\Phi^0_\SPT(1)
%\prod_{j=1}^n \Phi_\SP^j(s_j) \Phi_\IS^j(t_j) \right ] \right . \nonumber \\
%&\times& \left . \left[\Phi^0_\SPT(2)
%\prod_{j=n+1}^{2n} \Phi_\SP^j(s_j) \Phi_\IS^j(t_j) \right]\right. \nonumber \\
%&\times&\left . \chi_\bR^0(\bR)\, \chi_\CM^0(\bR_\CM) \right\},
%\end{eqnarray}
\begin{eqnarray}
\label{manysei}
\Psi = &{\cal A}&\left\{\left[\Phi^0_\SPT(1) \prod_{j=1}^n \Phi_\SP^j(s_j) \Phi_\IS^j(t_j) \right] \right . \nonumber \\
&\times& \left. \left[\Phi^0_\SPT(2) \prod_{j=n+1}^{2n} \Phi_\SP^j(s_j) \Phi_\IS^j(t_j) \right]
\,\chi_\bR^0(\bR)\, \chi_\CM^0(\bR_\CM) \right\},
\end{eqnarray}
where ${\cal A}$ indicates antisymmetrization and normalization, the functions $\Phi_\SP^j(s_j)$
and $\Phi_\IS^j(t_j)$ refer respectively to the spin $s_j$ and to the isospin $t_j$ of
the nucleons which compose the clusters, and, finally, $\chi_\bR^0(\bR)$ and $\chi_\CM^0(\bR_\CM)$
describe respectively the relative motion of the clusters and the motion of their center of mass. 
Then, we use the following identity \cite{Wildermuth2}:
\beq
\label{manysette}
\sum_{j=1}^{2n}r_j^2=\sum_{j=1}^n\left(r_j^\inter\right)^2+\sum_{j=n+1}^{2n}\left(r_j^\inter\right)^2 
+\frac{n}{2} \left\{ (\bR_1-\bR_2)^2+(\bR_1+\bR_2)^2\right\}.
\eeq
Therefore, if it is assumed that
\begin{eqnarray}
\label{n1}
\chi_\bR^0(\bR)&=&\exp \left(-\frac{m\omega}{2}\left\{\frac{n}{2}(\bR_1-\bR_2)^2\right\}\right)\,\chi(\bR),\nonumber \\
\label{n2}
\chi_\CM^0(\bR_\CM)&=&\exp \left(-\frac{m\omega}{2}\left\{\frac{n}{2}(\bR_1+\bR_2)^2\right\}\right)\, \chi_\CM(\bR_\CM), \nonumber
\end{eqnarray}
the wavefunction (\ref{manysei}) can be rewritten as follows:
\beq
\label{manyotto}
\Psi = {\cal A}\left\{\exp\left(-\frac{m\omega}{2}\sum_{j=1}^{2n} r_j^2\right)\, \chi(\bR)\,\chi_\CM(\bR_\CM)
\,\Phi_\ST(1) \Phi_\ST(2)\right\},
\eeq
where we pose
\begin{eqnarray}
&&\Phi_\ST(1) = \prod_{j=1}^n \Phi_\SP^j(s_j) \Phi_\IS^j(t_j), \label{manydiecia} \\
&&\Phi_\ST(2) = \prod_{j=n+1}^{2n} \Phi_\SP^j(s_j) \Phi_\IS^j(t_j). \label{manydiecib}
\end{eqnarray}
But the functions $\exp(-\frac{m\omega}{2}\sum_{j=1}^{2n}r_j^2)$ and $\chi_\CM(\bR_\CM)$
can be taken out of the antisymmetrization. 
Thus, we can write:
\beq
\label{manynove}
\Psi = \chi_\CM(\bR_\CM)\exp\left(-\frac{m\omega}{2}\sum_{j=1}^{2n} r_j^2\right)
\,{\cal A}\left\{\chi(\bR)\,\Phi_\ST(1)\Phi_\ST(2)\right \},
\eeq
Now, we can introduce the Hamiltonian $H$ acting on the relative motion wavefunction $\chi(\bR)$; $H$
can be written as a sum of three terms: $H=T+V_D+\frac{1}{2}\sum_{q=1}^N\sum_{p\neq q}^N V_{q,p}$ $(N=2n)$. 
The first term $T$ denotes the kinetic energy of the relative motion of the clusters, 
$V_D$ is the potential of the direct forces 
acting between the clusters, and the last term is the sum of the nucleon--nucleon gaussian potential
which plays an essential role in the antisymmetrization, as it will be explained below.
Note that the Coulomb potential is omitted in the Hamiltonian in view of the fact that we are interested in
nuclear effects, like resonances and antiresonances, and, accordingly, in nuclear phase--shifts and
scattering amplitudes.
\begin{remark}
\rm
In connection with the Coulomb subtraction it is worth noting that:
{\it i)} Due to the long range of the Coulomb forces, the exchange part of the Coulomb interaction
practically does not greatly influence the scattering wavefunction (see ref. \cite{Wildermuth2} and the references
quoted therein). For a more detailed analysis and for a numerical comparison between
the $\alpha$--$\alpha$ phase--shifts computed with and without the exact exchange Coulomb interaction 
the interested reader is referred to Appendix B of ref. \cite{Tang} 
(see, in particular, Fig. 21 of this Appendix). \\
{\it ii)} For the sake of preciseness, it must be distinguished between 
{\it quasinuclear} phase--shifts \cite{Regnier1,Regnier2} and purely {\it nuclear} phase--shifts, 
which are those related to the scattering between the same particles but without the Coulomb interaction. 
It has been shown \cite{Regnier1,Regnier2} that the {\it quasinuclear} phase--shifts
$\delta^*_\ell$ differ from the corresponding {\it nuclear} ones $\delta_\ell$ by quantities
of the order $c\delta_\ell$, where $c=ZZ'e^2/\hbar v$ (with standard meaning of symbols).
The value of $c$ can be quite large at low energy, but this fact is not relevant for our
subsequent analysis, and therefore we will neglect this factor in the following.
\end{remark}

In a very rough model we could assume that direct forces of harmonic oscillator type: {\it i.e.}, 
$V_D=(K/4N)\sum_{p,q=1}^N|\br_p-\br_q|^2$ are still present. 
If the strength constant $K$ is small, then this potential gives rise to a negligible interaction among 
the nucleons belonging to the same cluster, whereas the force acting among nucleons belonging to different 
clusters, and which are quite far apart, is relevant. Furthermore, if we rewrite $V_D$ in terms 
of hyperspherical coordinates, we have $V_D=K\rho^2/2$, and we can separate the radial from the angular 
variables in the equation of motion.
When we move back from the hyperspherical coordinates to the ordinary spatial ones $\br_j$, the potential
$V_D(\rho)$ yields two terms: one depending on the square of the modulus of the relative coordinate,
the other one depending on the center of mass of the clusters and on the totally symmetric
function of the coordinates $\sum_{j=1}^{2n} \br_j^2$ (see (\ref{manyquattro}) and (\ref{manysette})).
Therefore, we keep denoting (with a small abuse of notation) by $V_D=V_D(|\bR|)$ the potential corresponding
to the direct interaction of harmonic oscillator type acting on the relative motion wavefunction
$\chi(\bR)$. \\
When the two clusters penetrate each other, the effect of the direct forces decreases rapidly,
while other types of interactions between nucleons come into play. 
The nucleon--nucleon interaction that accounts for the exchange, and that is used in the
antisymmetrization process is generally represented by a potential of gaussian form \cite{Wildermuth2}:
$V_{p,q} \propto V_0 \exp (-K' |\br_p-\br_q|^2) \{w (1+P_{pq}^r)\}$, $P_{pq}^r$ being the operator
that exchanges the space coordinates of the $p$--th and $q$--th nucleons, and $w$ a constant.
A minimization of functionals, in the sense of Ritz variational calculus \cite{Wildermuth2}, in which
the nucleon--nucleon interaction is described by both harmonic and exchange
potentials of gaussian form, yields Euler--Lagrange equations which contain, in addition to potentials of the
form $V_D(|\vec{R}|)$, also non--local potentials of the form $V(\vec{R},\vec{R'})$ that
still preserve the rotational invariance in a sense that will be clarified below.
Working out the problem in this scheme, all the methods in use ({\it i.e.}, resonating group, 
complex generator coordinate and cluster coordinate methods) lead to an integro--differential 
equation of the form \cite{Wildermuth2}:
\beq
\label{manyundici}
\{-\Delta +V_D\}\chi(\bR)+g\int_{\R^3} V(\bR,\bR')\chi(\bR')\,d\bR'=E\chi(\bR),
\eeq
where $\hbar=2\mu=1$ ($\mu$ is the reduced mass of the clusters), 
$g$ is a real coupling constant, $E$, in the case of the scattering process, represents the 
scattering relative kinetic energy of the two clusters in the center of mass system, and $\Delta$ is the 
relative motion kinetic energy operator.
Finally, let us note that the integral in eq. (\ref{manyundici})
can include a local potential: {\it i.e.}, we can formally write $V(\bR,\bR')=V_E(\bR,\bR')+V_D(\bR,\bR')\delta(\bR-\bR')$,
where $V_E$ and $V_D$ represent the terms that derive from exchange and direct forces respectively, and
$\delta$ is the Dirac distribution.

If we want to develop from equation (\ref{manyundici}) a scattering theory which describes the cluster collision,
some additional conditions must be imposed. First, the current conservation law requires that
the current of the incoming particles is equal to the current of the outgoing particles. It follows that
$V(\bR,\bR')$ is a real and symmetric function: $V(\bR,\bR')=V^*(\bR,\bR')=V(\bR',\bR)$. 
Moreover, we remark once more that either the nucleon--nucleon potentials (which are of harmonic or gaussian type) 
and the wavefunctions are rotationally invariant. Then $V(\bR,\bR')$ depends
only on the lengths of the vectors $\bR$ and $\bR'$ and on the angle $\gamma$ between them, or
equivalently on the  dimension of the triangle with vertices $(0,\bR,\bR')$ but 
not on its orientation. Hence, $V(\bR,\bR')$ can be formally expanded as follows:
\beq
\label{manyundicibis}
V(\bR,\bR')=\frac{1}{4\pi R R'}\sum_{s=0}^\infty (2s+1) V_s(R,R') P_s(\cos\gamma),
\eeq
where $\cos\gamma=(\bR\cdot\bR')/(R R')$, and $P_s$ are the Legendre polynomials.
The Fourier--Legendre coefficients $V_s(R,R')$ are given by:
\beq
\label{manyundicitris}
V_s(R,R')=4\pi R R' \int_{-1}^1 V(\bR,\bR'; \cos\gamma) P_s(\cos\gamma)\, d(\cos\gamma).
\eeq
We may therefore conclude that the l.h.s operator of eq. (\ref{manyundici}), acting on the function
$\chi$, is a formally hermitian and rotationally invariant operator.

Next, we expand the relative motion wavefunction $\chi(\bR)$ in the form:
\beq
\label{manyundiciquater}
\chi(\bR) = \frac{1}{R} \sum_{\ell=0}^\infty \chi_\ell(R) \, P_\ell(\cos\theta),
\eeq
where now $\ell$ is the relative angular momentum between the clusters.

Since $\gamma$ is the angle between the two vectors $\bR$ and $\bR'$, whose directions are determined
by the angles $(\theta,\phi)$ and $(\theta',\phi')$ respectively, we have:
$\cos\gamma=\cos\theta\cos\theta' + \sin\theta\sin\theta' \cos(\phi-\phi')$. Then, the following addition
formula for the Legendre polynomials can be stated:
\beq
\label{manyundiciquinquer}
\int_0^\pi \int_0^{2\pi} P_s(\cos\gamma)P_\ell(\cos\theta')\sin\theta'\, d\theta'\, d\phi'
= \frac{4\pi}{(2\ell+1)}P_\ell(\cos\theta)\delta_{s\ell}\,.
\eeq
By substituting expansions (\ref{manyundicibis}) and (\ref{manyundiciquater}) in (\ref{manyundici}),
and taking into account (\ref{manyundiciquinquer}), we obtain:
\beq
\label{manydiciassette}
\chi''_\ell(R)+k^2 \chi_\ell(R) -\frac{\ell(\ell+1)}{R^2} \chi_\ell(R)
= g\int_0^{+\infty} V_\ell(R,R') \chi_\ell(R') \, dR',
\eeq
where $k^2=E$; the local potential, which is supposed to be included in the non--local one, has been omitted.

To carry the analysis a step forward, we impose a bound on the potential which will turn out to be very useful
later on (see, in particular, the norm of the Hilbert space defined by formula (\ref{manyventuno})). 
We suppose that the function $V(\bR,\bR')$ is a measurable function in $\R^3 \times \R^3$, 
and we also assume there exists a constant $\alpha > 0$ such that:
\beq
\label{manymany}
C = \left\{\int_{\R^3} (1+R^2)e^{2\alpha R}\,d\bR \int_{\R^3}(1+R'^2) R'^2 e^{2\alpha R'}\,
|V(\bR,\bR')|^2 \, d\bR' \right\}^{1/2} < \infty.
\eeq
Let us note that bound (\ref{manymany}) restricts the class of potentials admitted for what concerns
the order of the singularities at the origin and the growth properties at infinity.
If bound (\ref{manymany}) is satisfied, then expansion
(\ref{manyundicibis}) converges in the norm $L^2(-1,1)$ for almost every $R$, $R' \in [0,+\infty)$.
If we substitute expansion (\ref{manyundicibis}) into equality (\ref{manymany}), and integrate
with respect to the angular variables, from the Parseval equality we get:
\beq
\label{manymanybis}
C = \left\{\int_0^{+\infty}(1+R^2)e^{2\alpha R}\,dR\,
\int_0^{+\infty}(1+R'^2) R'^2 e^{2\alpha R'}\,
\left (\sum_{s=0}^\infty(2s+1)V_s^2(R,R')\right )\,dR'\right\}^{1/2},
\eeq
and, consequently, $V_\ell(R,R')$ must necessarily satisfy the following condition:
\beq
\label{manymanytris}
C_\ell = \left\{\int_0^{+\infty}(1+R^2)e^{2\alpha R}\,dR\,
\int_0^{+\infty}(1+R'^2) R'^2 e^{2\alpha R'}\,
V_\ell^2(R,R')\,dR'\right\}^{1/2} < \frac{C}{(2\ell+1)},
\eeq
which represents a constraint on the $\ell$--dependence of $V_\ell(R,R')$.
\begin{remark}
\rm
At this point we want to note:
{\it i)} From condition (\ref{manymanytris}) it derives that the lifetimes of the rotational resonances decrease
for increasing values of $\ell$, in agreement with the phenomenological data (see the analysis which
follows from next formula (\ref{manytrenta})). \\
{\it ii)} Bounds (\ref{manymany})--(\ref{manymanytris}) do not admit, for instance, direct potentials of the form
$V_D(|\vec{R}|)\propto R^2$. This difficulty can be overcome by a suitable modification of the
shape of the potential at large values of $R$: {\it i.e.}, imposing an exponential tail for
$R>R_0$ ($R_0$ being a constant). This modification, which refers exclusively to the Hamiltonian acting
on the relative motion wavefunction $\chi(\vec{R})$, may be regarded as a small perturbation
(if $R_0$ is sufficiently large), which is irrelevant in connection with the group theoretical analysis
of the spectrum previously performed.
\end{remark}

Now, we must distinguish between two kinds of solutions of eq. (\ref{manydiciassette}): the scattering solutions
$\chi_\ell^s(k,R)$, and the {\it bound state} solutions $\chi_\ell^b(R)$. 
\begin{itemize}
\item[i)] The scattering solutions satisfy the condition:
\begin{eqnarray}
&&\chi_\ell^s(k,R)=kRj_\ell(kR)+\Phi_\ell(k,R), \label{manydiciottoa}\nonumber \\
&&\sds\Phi_\ell(k,0)=0,~
\lim_{R\rightarrow +\infty}\left\{\frac{d}{dR}\Phi_\ell(k,R)-ik\Phi_\ell(k,R)\right\}=0, \label{manydiciottob} \nonumber
\end{eqnarray}
where $j_\ell(kR)$ are the spherical Bessel functions, and the functions $d\Phi_\ell/dR$
are supposed to be absolutely continuous.
\item[ii)]
The {\it bound state} solutions $\chi_\ell^b(R)$ satisfy the condition:
\beq
\label{manydiciannove}
\int_0^{+\infty} \left | \chi_\ell^b(R) \right |^2 \, dR < \infty~,~~~~~\chi_\ell^b(0)=0.
\eeq
\end{itemize}

\smallspace
\noindent
The problem of solving the integro--differential equation (\ref{manydiciassette}), with conditions
(i) or (ii) can be reduced to the problem of solving the linear integral equation of the Lippmann--Schwinger
type \cite{Bertero1,Bertero2}:
\beq
\label{manyventi}
v_\ell(k,R)=v_{\ell,0}(k,R)+g\int_0^{+\infty}L_\ell(k;R,R')v_\ell(k,R')\,dR',
\eeq
where
\begin{eqnarray}
v_{\ell,0}(k;R)&=&\int_0^{+\infty}k R'\, V_\ell(R,R')\, j_\ell(kR')\,dR', \label{manyventia}~~~~~~~~~~~~~ \\
L_\ell(k;R,R')&=&\int_0^{+\infty}V_\ell(R,t)G_\ell(k;t,R')\, dt, \label{manyventib} \\
G_\ell(k;t,R') &=& -iktR'\, j_\ell(k\min\{t,R'\})\, h_\ell^{(1)}(k\max\{t,R'\}), \label{manyventic}
\end{eqnarray}
$h_\ell^{(1)}$ denoting the spherical Hankel functions.

It is convenient to rewrite eq. (\ref{manyventi}) as a linear equation in a suitable functional space
$X$. Let us introduce the Hilbert space \cite{Bertero2}:
\beq
\label{manyventuno}
X = \left\{x(\bR) \, : \|x\|_X = 
\left [\int_0^{+\infty}\left ( 1+R^2\right )e^{2\alpha R} |x(R)|^2\, dR\right ]^{1/2} < +\infty \right\},
\eeq
with inner product
\beq
\label{manyventidue}
(x,y)_X = \int_0^{+\infty} \left ( 1+R^2\right )e^{2\alpha R} x(R) y^*(R) dR~,~~~~~(x,y\in X).
\eeq
Then eq. (\ref{manyventi}) can be rewritten as
\beq
\label{manyventitre}
\left [ 1-gL_\ell(k)\right ] v_\ell(k,\cdot) = v_{\ell,0}(k,\cdot).
\eeq
In refs. \cite{Bertero1,Bertero2} and in the Appendix it is proved that
for any $k$ in the half--plane $\Imag k \geq -\alpha$
$(\alpha > 0)$, the operator $L_\ell(k)$ is compact on $X$, and, therefore, the Fredholm alternative
applies to eq. (\ref{manyventitre}) if $v_{\ell,0}(k,\cdot) \in X$.
The latter condition is satisfied for any $k$ in the strip $|\Imag k | \leq \alpha$ $(\alpha >0)$,
provided that bound (\ref{manymany}) is satisfied. Then, from the Fredholm alternative,
it follows that either there exists in $X$ (for $|\Imag k | \leq \alpha$) a non--trivial solution of the
homogeneous equation:
\beq
\label{manyventiquattro}
\left [ 1-gL_\ell(k)\right ] v_\ell(k,\cdot) = 0,
\eeq
or a solution in $X$ ($|\Imag k | \leq \alpha$) of eq. (\ref{manyventitre}) exists, and is unique. Besides,
the map $k \rightarrow L_\ell(k)$ is an operator--valued function holomorphic in the half--plane
$\Imag k \geq -\alpha$ $(\ell=0,1,2,\ldots)$, and the map $k \rightarrow v_{\ell,0}(k,\cdot)$ is a holomorphic
vector--valued function in the strip $|\Imag k | < \alpha$ $(\ell=0,1,2,\ldots)$.
Therefore, for positive and real values of $R$, the scattering solution $\chi_\ell^s(k,R)$
is holomorphic in the strip $|\Imag k | < \alpha$, except at those $k$--points where a non--zero solution
of the homogeneous eq. (\ref{manyventiquattro}) exists.

\begin{figure}[ht]
\centering
\includegraphics[width=3.2in]{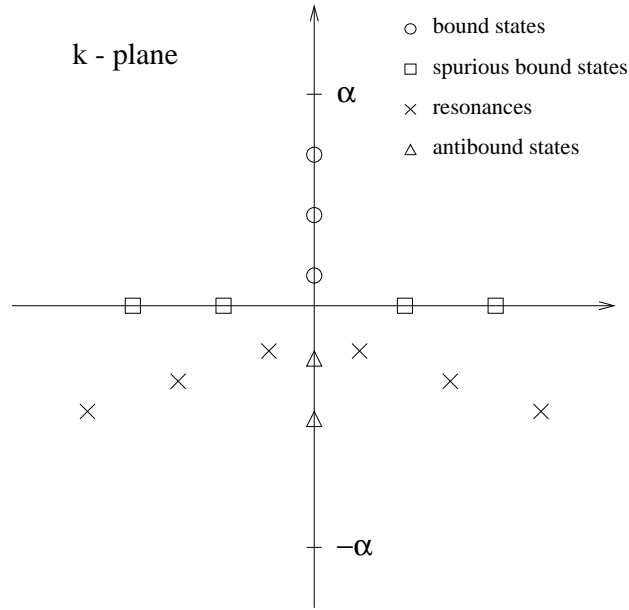}
\caption{Representation of bound states and resonances in the $k$--plane.}
\label{figura_1}
\end{figure}

As in the case of local potentials, one can compare the asymptotic behaviour of the scattering solution,
for large values of $R$, with the asymptotic behaviour of the free radial function $j_\ell(kR)$, and, correspondingly,
introduce the phase shifts $\delta_\ell(k)$. Accordingly, one can then define the scattering amplitude
\beq
\label{manyventicinque}
t_\ell(k) = e^{i\delta_\ell(k)}\,\sin\delta_\ell(k),
\eeq
and prove that $t_\ell(k)$ has the same analyticity domain as $\chi_\ell^s(k,\cdot)$.
Finally, the following asymptotic behaviour of the phase--shifts, for $\ell\rightarrow\infty$, can be proved 
\cite{Bertero2}:
\beq
\label{manyventisei}
\delta_\ell(k) = O\left (\ell^{-1}\, e^{-\beta \ell}\right ),~~\cosh\beta = 1+\frac{2\alpha^2}{k^2},
~~(\ell\rightarrow\infty).
\eeq
On the other hand, if a non--zero solution of the homogeneous equation (\ref{manyventiquattro}) exists,
we then have a singularity of the resolvent $R_\ell(k,g)$, which reads:
\beq
\label{manyventisette}
R_\ell(k,g) = \left [ 1-gL_\ell(k)\right ]^{-1},
\eeq
and several cases occur. The operator--valued function $k\rightarrow R_\ell(k,g)$ is
meromorphic in $\Imag k > - \alpha$, and we may associate a precise physical meaning to its
singularities, which are isolated poles.
First of all, observe that, since the coupling constant $g$ is real and
$V(\bR,\bR')=V^*(\bR,\bR')=V(\bR',\bR)$, the poles of $R_\ell(k,g)$ (at fixed $g$) which lie in the
half--plane $\Imag k \geq 0$ can occur only for $\Real k = 0$, or for $\Imag k \geq 0$. Then,
we must consider four cases (see fig. \ref{figura_1} for a graphical presentation):
\begin{itemize}
\item[a)]
The poles of $R_\ell(k,g)$ that lie on the imaginary axis at $k=ib$ $(b>0)$; they correspond to bound states of energy
$E=-b^2$.
\item[b)]
The poles of $R_\ell(k,g)$ that lie on the real axis, {\it i.e.}, at $k=b$ ($b$ real); they correspond to
{\it spurious bound states} of energy $E=b^2$.
These poles are distributed in pairs symmetric with respect to $k=0$.
\item[c)] The poles of $R_\ell(k,g)$ that lie in the strip $-\alpha \leq \Imag k < 0~(\alpha > 0)$; they
are isolated, and may be interpreted as resonances if $\Real k \neq 0$. They occur in pairs symmetric
with respect to the imaginary axis (see fig. \ref{figura_1}).
\item [d)] The poles of $R_\ell(k,g)$ that lie on the imaginary axis at $k = -ib$ $(b>0)$; they correspond
to {\it antibound states}. The wavefunctions corresponding to the antibound states do not
belong to $L^2[0,+\infty)$. These states show up on the low--energy behaviour of the cross--section
if the binding energy of the state is sufficiently small. It is, in general, difficult to attach
any relevant physical meaning to antibound states as one usually does for the bound states
or with the resonances if their width is small \cite{DeAlfaro}. Therefore, we shall not 
deal with them again.
\end{itemize}

\begin{remark}
\rm
The only observable quantities are bound states and
cross sections. From the latter one can derive the physical phase--shifts $\delta_\ell(k)$ (with $k$
real and non--negative), which can still be regarded as {\it measurable quantities}. Therefore, the
half--axis $\Real k \geq 0$ is usually called ``physical region''. The analytical continuation from
the physical region to the complex $k$--plane is, however, of great importance since the poles of the
resolvent appear as bound states or resonances; the latter are observed as peaks in the cross section.
\end{remark}

From the viewpoint of our analysis, it deserves some interest an inequality which holds true for any real or
imaginary value of $k$ \cite{Bertero2}:
\beq
\label{manytrenta}
\left \| L_\ell(k)\right \|_X = \sup_{x\in X} \frac{\|L_\ell(k) x\|_X}{\|x\|_X}\leq\frac{1}{2}
\pi^{3/2} \frac{C}{(2\ell+1)}.
\eeq
It follows that, if we set $L=\frac{1}{2}(\frac{1}{2}|g|\pi^{3/2}C-1)$, for $\ell>L$, $gL_\ell(k)$ is a
contraction in $X$, and, therefore, for $\ell>L$ no bound state (corresponding to imaginary values of $k$)
or {\it spurious bound state} solutions (corresponding to real values of $k$) can exist. Let us now focus
our attention on the {\it spurious bound states}; inequality (\ref{manytrenta}) means that, for sufficiently
large $\ell$, the potentials $V_\ell$ are not strong enough to allow the existence of bound states embedded in the
continuum. If, however, we add to $k$ a term $-ib$ $(b>0)$, constraint (\ref{manytrenta}) 
no longer holds true, 
and we can have poles in the lower half--plane ({\it i.e.}, $\Imag k < 0$), corresponding to resonances whose lifetime 
is related to $b$ (remember that the {\it spurious bound state} poles are
distributed in pairs symmetric with respect to $k=0$, similarly to the singularities corresponding to
the resonances which are symmetrically distributed with respect to the imaginary axis (see fig. \ref{figura_1})).
For increasing values of $\ell$ the r.h.s. of bound (\ref{manytrenta}) becomes smaller, and, correspondingly,
the admitted potentials $V_\ell$ become weaker (see also bound (\ref{manymanytris})); accordingly, they cannot sustain
the trapping which generates the resonances for a long time. The lifetime of the resonances becomes shorter for 
increasing values of the angular momentum in agreement with the spectrum of the rotational
bands of resonances:
the latter evolve into surface waves. In this case, we move from quantum to semiclassical phenomena that
cannot be properly described using spectral theory: the surface waves cannot be regarded as
unstable states.

Reverting to the phase--shifts $\delta_\ell(k)$, let us note that the
resonance poles in the $k$--plane necessarily contain an imaginary part which is related to the
resonance lifetime. Therefore, we can always guarantee the existence of a scattering solution, and,
consequently, of the associated phase--shift, for any physically measurable value of energy arbitrarily
close to the resonances. Two remarkable features of the $\delta_\ell(k)$ behaviour are worth being mentioned: 
\begin{itemize}
\item[i)] 
If $\delta_\ell(k)$ is supposed to be close to zero, and below the resonance,
then its value will increase passing through $\pi/2$ just when the energy crosses the energy location
of the resonance. Accordingly, we have $\sin^2\delta_\ell=1$ at the resonance energy, and the cross section
will show a sharp maximum. 
\item[ii)] 
In view of the asymptotic behaviour of $\delta_\ell(k)$, for $\ell\rightarrow\infty$ (see 
(\ref{manyventisei})), we have $\delta_\ell(+\infty)=0$. Therefore, after an increase due to a resonance,
$\delta_\ell(k)$ will necessarily pass downward through $\pi/2$. Correspondingly, we have an
antiresonance or an {\it echo}. 
\end{itemize}
The width of a resonance measures (inversely) the time delay of a scattered wave packet due to the
trapping of the incoming cluster. This process can be viewed as a collective phenomenon, and it
will be described in detail in the next section. On the contrary, there is no trapping at an echo
energy, and its width measures the time advance of the packet. The echoes are due to the 
repulsive forces which derive from the exchange effects and from the antisymmetrization. 

\section{Collective Coordinates and Hydrodynamical Model of the Trapping: Spin--Width of the Rotational Resonances}
\label{se:collective}
This section is devoted to an analysis of the rotational resonances, regarded as a
collective phenomenon. Therefore, it is necessary to introduce appropriate coordinates that
make it possible to separate the collective from the single particle dynamics.
Zickendrath \cite{Zickendrath1,Zickendrath2,Zickendrath3}
proposed such a type of coordinates (see also ref. \cite{Gallina}), hereafter called $Z$--coordinates. 
We first illustrate the passage from Jacobi to $Z$--coordinates in 
the simple case of the three--body problem, then the procedure will be generalized to $N$ particles. 
Let the system be described by two Jacobi coordinates $\bxi_1$, $\bxi_2$; we can introduce
a ``kinematic rotation'' in the sense of Smith \cite{Smith}, and replace $\bxi_1$, $\bxi_2$ by the vectors
$\by_1$, $\by_2$ obtained as follows:
\begin{eqnarray}
\by_1 &=& \bxi_1 \, \cos\eta + \bxi_2 \, \sin\eta, \label{rounoa} \\
\by_2 &=& -\bxi_1 \, \sin\eta + \bxi_2 \, \cos\eta. \label{rounob}
\end{eqnarray}
Then, we look for the value $\eta_0$ of $\eta$ such that the vectors $\by_1$ and $\by_2$ are
orthogonal: $\by_1\cdot\by_2 = 0$. We thus obtain:
\beq
\label{rotre}
\eta_0 = \frac{1}{2}\tan^{-1}\frac{2\bxi_1\cdot\bxi_2}{\bxi_1^2-\bxi_2^2}.
\eeq
It can be shown \cite{Zickendrath2} that the directions of $\by_1$ and $\by_2$,
obtained by the kinematic rotation (\ref{rounoa}, \ref{rounob}) with $\eta = \eta_0$,
coincide with the principal axes of the moment of inertia in the plane of the three particles.
We can then consider the Euler angles $\phi,\theta,\psi$ of the three axis $\by_1$, $\by_2$, and
$\by_1 \wedge \by_2$ in the center of mass system, and, finally, replace the Jacobi coordinates
$\bxi_1$, $\bxi_2$ by: $\phi,\theta,\psi;|\by_1|,|\by_2|,\eta_0$. \\
Now consider an arbitrary number $N$ of particles of equal mass $m$; by extending formulae 
(\ref{rounoa}, \ref{rounob}),
we write:
\beq
\label{roquattro}
\by_i = \sum_{k=1}^{N-1} a_{ki} \bxi_k\,,~~~~~(i=1,2,3),
\eeq
where the coefficients $a_{ki}$ are elements of an orthogonal matrix. Since for orthogonal matrices
inverse and transposed matrix coincide, system (\ref{roquattro}) can be easily inverted:
\beq
\label{roquattrouno}
\bxi_i = \sum_{k=1}^3 a_{ik} \by_k\,,~~~~~(i=1,2,\ldots,N-1).
\eeq
Furthermore, the orthogonality conditions give:
\beq
\label{roquattrodue}
\sum_{i=1}^{N-1} a_{ik} a_{ij} = \delta_{kj}\,,~~~~~(k,j=1,2,3).
\eeq
In addition, we require that the vectors $\by_1$, $\by_2$, $\by_3$ are perpendicular to each other: {\it i.e.},
\beq
\label{roquattrotre}
\by_k \cdot \by_j = |\by_k| \, |\by_j| \, \delta_{kj}\,.
\eeq
In conclusion, the Jacobi vectors $\bxi_1,\ldots,\bxi_{N-1}$ can be replaced by the following
coordinates:
\begin{itemize}
\item[i)]
the lengths of the vectors $\by_1,\by_2,\by_3$, which are perpendicular to each other and directed
along the principal axes of the inertia ellipsoid;
\item[ii)]
the Euler angles $\phi,\theta,\psi$ which describe the positions of the three axes
$\by_1,\by_2,\by_3$ in the center of mass system;
\item[iii)]
the coefficients $\{a_{ik}\}$ of system (\ref{roquattrouno}), constrained by conditions
(\ref{roquattrodue}), which can be regarded as {\it internal coordinates}.
\end{itemize}
If we assume that the colliding clusters have spherical shape, and, in addition, that
they are composed of an equal number of particles, each of mass $m$, then the interaction model 
proposed by Zickendrath \cite{Zickendrath3} for the $\alpha$--$\alpha$ elastic scattering can be easily 
generalized. We can observe that, in this model, the direction of vector $\bR$ describing the 
relative coordinate between the clusters coincides with the direction of one of the vectors $\by_i$. 
Therefore, instead of using the relative coordinate $\bR$, it is more convenient to describe the relative motion
of the clusters with a vector whose direction and length are $\theta,\phi$ and $|\by|$. Then the
wavefunction that we want to consider will be
\beq
\label{rosedici}
\Psi = \chi(\by) {\cal A} \left\{\Phi(1)\Phi(2)\right\},
\eeq
where $\Phi(1)$ and $\Phi(2)$ are the wavefunctions that describe the clusters 1 and 2, respectively.
We have thus factorized, through formula (\ref{rosedici}), the wavefunction into the products of two factors:
one depending only on the collective coordinates $|\by|,\theta,\phi$, and the other depending only on the 
internal coordinates (compare (\ref{rosedici}) with (\ref{manynove})). 
Now remember that the resonances being considered are produced
by the rotation of the clusters around their center of mass, and, in this process, the antisymmetrization
of the fermions belonging to different clusters, can be neglected in a first rough approximation. 
Therefore we are essentially concerned with only the 
function $\chi(\by)$. Moreover, if we suppose that the energy is high enough to allow a semiclassical
approximation, then $\chi(\by)$ can be written in the following form: $\chi=A\exp(i\Theta)/\sqrt{2}$, and, 
accordingly, the current density reads
$\bj = i\{\chi\nabla\chi^*-\chi^*\nabla\chi\}=A^2\nabla\Theta$. 
In this way we may introduce a velocity field, and regard $\Theta$ as a velocity
potential in the hypothesis of irrotational flow, {\it i.e.}, $\bv = \nabla\Theta$.

\begin{figure}[htb]
\centering
\includegraphics[width=3.0in]{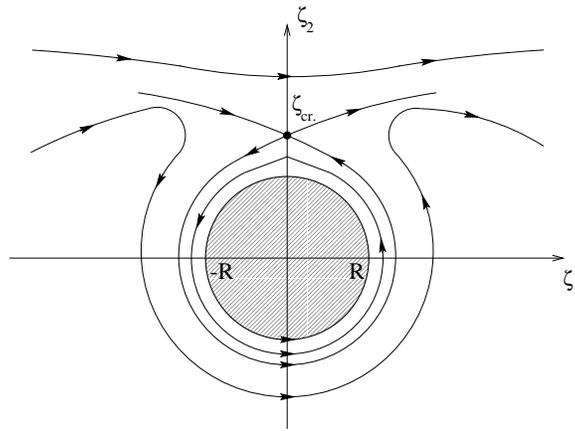}
\caption{Hydrodynamical picture of the trapping.}
\label{figura_2}
\end{figure}

At this point we have all that is needed to present a hydrodynamical picture of the trapping which is able to
produce rotational resonances. 
In order to construct this hydrodynamical model, it is more suitable to describe
the process in the laboratory frame, and represent the incoming beam as a flow streaming around the target.
We then work out our model in a plane using only two coordinates: the radial coordinate
and the angle $\theta$ (the angle $\phi$ can be ignored as explained in the remark below).
Representing the velocity field in the complex $\zeta$--plane, we denote by $f(\zeta)$,
($\zeta=\zeta_1+i\zeta_2$) the complex potential of the flow.
First, we begin with an irrotational flow around the circle $C_R$ ($R$ is the radius of the circle), whose
corresponding complex potential has the form: $f(\zeta)=\bv_\infty (\zeta+R^2/\zeta)$, where $\bv_\infty$ 
is the flow velocity at infinity, chosen to be parallel to the $\zeta_1$--axis. 
At the points of the circle $C_R$, which is a streamline, the velocity
is directed tangentially to the circle, and vanishes at two critical points: $\zeta=-R$, where
the streamline branches into two streamlines coinciding with the upper and lower semicircles of $|\zeta|=R$,
and at $\zeta=R$, where these streamlines converge again into the single straight line $\zeta_2=0$.
Now, let us add a vortex term of the form: $(\Gamma_\infty/2\pi i)\ln\zeta$ to give the whole potential flow:
\beq
\label{rodiciassette}
f(\zeta)=\bv_\infty\left (\zeta + \frac{R^2}{\zeta}\right ) + \frac{\Gamma_\infty}{2\pi i}\ln\zeta,
\eeq
where $\Gamma_\infty$ is the vortex strength
\beq
\label{rodiciotto}
\Gamma_\infty = \oint_{C_R} f'(\zeta)\, d\zeta\,.
\eeq
The critical points are now given by
\beq
\label{rodiciannove}
\zeta_{cr.} = i \frac{\Gamma_\infty}{4\pi\bv_\infty}\pm\left (R^2-\frac{\Gamma_\infty^2}{16\pi^2\bv_\infty^2}
\right )^{1/2}.
\eeq
When $\left | \Gamma_\infty/4\pi\bv_\infty \right | > R$, in the domain $|\zeta | > R$ there is only one
critical point lying on the imaginary $\zeta_2$--axis. Through this point
passes the streamline that separates the closed streamlines of the flow from the open streamlines 
(see fig. \ref{figura_2}). Thus, we have obtained the trapping produced by the vortex. 
Note that for increasing values of $|\bv_\infty|$ (at fixed $\Gamma_\infty$) the inequality
$\left | \Gamma_\infty/4\pi\bv_\infty\right | > R$ ceases to hold, and, accordingly,
no trapping is allowed. In conclusion, the resonance can be heuristically depicted
as a vortex, and, accordingly, the rotational flow produces a vorticity ${\bf \omega} = \nabla \times \bv$.

\begin{remark} 
\rm
In the representation of this hydrodynamical model of the trapping we are forced
to choose an orientation of the vortex (see the counterclockwise orientation in fig. \ref{figura_2}).
However, note that this orientation is irrelevant since it corresponds to the 
determination of a phase factor that depends on $\phi$, which is not observable at quantum level,
in the absence of an appropriate external perturbation.
\end{remark} 

We are thus naturally led to define the spin--width of the resonance through the uncertainty principle
for the angular momentum. This is a very delicate question that has given rise to extensive 
literature \cite{Biedenharn}.
In fact, no self--adjoint operator exists with all the desiderable properties for an acceptable quantum 
description of an angular coordinate. If we try to write, in the more conventional form, the
standard dispersion inequalities we are led to the paradoxical situation of having an infinite spread in angles
for states sharp in angular momentum, while the physical meaning of the angle restricts its values 
to a finite range.
However, this difficulty can be overcome by introducing the exponentials of the angle variables. We proceed
as follows: first of all, we fix the canonical variables which come into play. In our case they are:
the angular momentum vector $\bL$ and the canonical angle conjugate to $(\bL^2)^{1/2}$, {\it i.e.},
the angle swept out in the orbital plane.
\begin{remark} 
\rm
The use of the $Z$--coordinates allows us to separate the {\it external} orbital angular
momentum $\bL_e$ from the {\it internal} orbital angular momentum $\bL_i$. In the approximation of our model, 
we can assume that the internal part of the wavefunction of each cluster is an eigenfunction of $\bL^2_i$ 
with null eigenvalue. This is a reasonable approximation in the assumption of spherical clusters, and if we
suppose that the tensor forces are of no great relevance. Therefore, the ground state of each cluster 
can be viewed approximately as an eigenfunction of $\bL^2_i$ with null eigenvalue. If such approximation 
holds true, then we only remain with the external orbital angular momentum which can be identified with 
the vector $\bL$ conjugate to the orbital angle in the sense explained above.
\end{remark} 

Next, we consider the exponential of the angle $Q = \theta+\frac{\pi}{2}$, {\it i.e.}, $e^{iQ}$, and the operator
$\bP = [(\bL^2+\frac{1}{4})^{1/2}-\frac{1}{2}]$. Then the minimum uncertainty in the dispersions
$\Delta\bP$ and $\Delta(\cos Q)$ is given by (see ref. \cite{Biedenharn}):
\beq
\label{roventi}
\Delta\bP \Delta(\cos Q) = \frac{1}{2} < \sin Q >.
\eeq
Now consider states that have a sharp value of the observable operator $\cos Q$.
The dispersion $\Delta(\cos Q)$ vanishes, and the expectation value
$< \sin Q >$ has a finite value. Hence, from the uncertainty relation (\ref{roventi}) the dispersion in
the angular momentum $\Delta\bP$ becomes unlimitedly large. This behaviour is exactly as would be expected 
on physical grounds. A maximal sharp {\it angle} observable implies a maximally spread value of the angular 
momentum.

Coming back to our physical problem, we can say that the resonances have a finite lifetime which corresponds
to the time of the trapping; after that, the unstable state decays. Then, we can speak of an 
{\it angular lifetime} of the resonance \cite{DeAlfaro} which gives the dispersion in the angle; 
correspondingly, we shall have a dispersion in the angular momentum as prescribed by the uncertainty 
relation (\ref{roventi}). We thus speak of a spin--width proper of the unstable states, 
which tends to zero as the angular lifetime tends to infinity. The bound states are, indeed, sharp 
in the angular momentum.

As explained in section \ref{se:many}, the poles of the resolvent $R_\ell(k,g)$ that correspond to the
resonances lie in the half--plane $\Imag k < 0$ (see fig. \ref{figura_1}), and their imaginary part
is related to the width of the resonance, which is inversely proportional to the time delay.
Analogously, we can represent the spin--width of the unstable states by extending the angular momentum
to complex values: the dispersion in the angular momentum, prescribed by the uncertainty relation
(\ref{roventi}), will be represented by the imaginary part of the angular momentum. The latter will be
denoted by $\lambda=\alpha+i\beta$. Since the angular momentum is complex, 
the centrifugal energy is complex too.
Neglecting the $\lambda$--dependence proper of the non--local interaction
we can write the continuity equation in the following form:
\beq
\label{roventuno}
\frac{\partial w}{\partial t} + \nabla\cdot\bj = 2 w \,\Imag \left < \frac{\lambda(\lambda+1)}{2\mu R^2}\right >,
~~~~~(\hbar = 1),
\eeq
where $\bj$ is the current density (already introduced above), $w = \chi^*\chi$, $\mu$ is the reduced mass, and
$R$ is the relative distance between the clusters. Then, we have:
\beq
\label{roventidue}
\Imag\left <\frac{\lambda(\lambda+1)}{2\mu R^2} \right > = 
\beta (2\alpha +1)\frac{1}{<2\mu R^2>} = \frac{\Gamma}{2},
\eeq
where $\Gamma$ is the width of the resonance. From (\ref{roventidue}) we get:
\beq
\label{roventitre}
\Gamma = \frac{\beta (2\alpha + 1)}{I},
\eeq
where $I=<\mu R^2 >$ is the moment of inertia of the system of clusters, regarded as a rigid rotator. 
This formula indicates that the values of $\Gamma$ increase for increasing values of 
$\alpha = \Real\lambda$, in perfect agreement with the phenomenological data \cite{Okai,Langanke}. 
This can easily be understood if we observe that, according to the result obtained at the end 
of section \ref{se:many}, the potentials $V_\ell$ become weaker for increasing values of $\ell$, and
therefore they cannot sustain the trapping proper of the resonance for a long time. 
This agrees with the hydrodynamical model, whose condition for producing the trapping
({\it i.e.}, $\left | \Gamma_\infty/4\pi\bv_\infty\right | > R$) suggests that at high energies
({\it i.e.}, high values of $|\bv_\infty|$, $\Gamma_\infty$ fixed), the trapping is not allowed. In conclusion,
$\beta$ can be regarded as the spin--width of the resonance, and its value increases for 
increasing energy.

This hydrodynamical model, and consequently the {\it spin--width} of the resonances, calls for the introduction
of the complex angular momentum plane into the description of the rotational band. Describing
the resonances by poles moving in the complex plane of the angular momentum (instead of using fixed poles)
allows to recover the global character of the rotational bands: 
{\it i.e.}, the grouping of resonances in families.
This latter method has been successfully used by one of us [GAV] in the phenomenological fits
of $\alpha$--$\alpha$, $\alpha$--$^{40}$Ca and $\pi^+$--p elastic scattering \cite{Viano1,Viano2,Fioravanti}.

Finally, let us note that this approach, giving an increase rate of the resonance width at higher
energy, explain the evolution of the rotational resonances into surface waves produced by diffraction,
even though at these energies the scenario is quite different, the inelastic scattering and the reaction
channels being dominant. At the present time diffraction phenomena and (nuclear and Coulomb)
rainbow mainly attract the theoretical and phenomenological attention 
(see, {\it e.g.}, refs. \cite{Nussenzveig,Khoa1,Khoa2,vonOertzen}). However, it is
one of the purposes of this paper to show that a deeper understanding of the evolution,
and, accordingly, of the global character of the (low energy) rotational bands can shed light
on these high energy phenomena.

\section*{Appendix}
\label{se:appendixb}
The results of the spectral analysis reported in section \ref{se:many} have been completely proved for
the case $\ell=0$ in ref. \cite{Bertero1}, and then partially extended to every integer value
of $\ell$ in ref. \cite{Bertero2}. This extension is complete if we observe that the proofs for the
case $\ell=0$ are based on the following bounds:
\begin{eqnarray}
|k R j_0(kR)| &=& |\sin kR | \leq \frac{2|k|R}{1+|k|R}e^{\alpha r},~~(|\Imag k |\leq\alpha),\nonumber \label{A1}\\
|G_0(k;R,R')| &\leq& \frac{2R}{1+|k|R} \, e^{\alpha(R+R')},\:~~(\Imag k \geq -\alpha),\nonumber \label{A2}
\end{eqnarray}
which can easily be extended to any integer value of $\ell$. 
For the spherical Bessel and Hankel functions the following majorizations hold true \cite{Newton}:
\begin{eqnarray}
|k R j_\ell(kR)| &\leq& \Cons \left [\frac{|k|R}{1+|k|R}\right ]^{(\ell+1)} \, e^{R\,|\SImag k |},~~~~ \label{A3}\\
|k R h_\ell^{(1)}(kR)| &\leq& \Cons \left [\frac{1+|k|R}{|k|R}\right ]^\ell \, e^{-R\,\SImag k}, \label{A4}
\end{eqnarray}
the constants depending only on $\ell$. Finally, from (\ref{manyventic}), (\ref{A3}) and (\ref{A4}) one gets:
\begin{eqnarray}
|k R j_\ell(kR)| &\leq& \Cons \frac{|k|R}{1+|k|R}\, e^{\alpha R},~|\Imag k| \leq \alpha, ~~~~~\label{A5}\nonumber \\
|G_\ell(k; R,R')| &\leq& \Cons \frac{R}{1+|k|R}\, e^{\alpha(R+R')},~ \Imag k \geq -\alpha,~~~~ \label{A6}\nonumber
\end{eqnarray}
and these bounds are sufficient for a complete generalization of the spectral results to any integer $\ell$.

\end{document}